\pdfoutput=1
\documentclass[12pt,a4paper]{article}

\usepackage[T1]{fontenc}
\usepackage{lmodern}
\usepackage{amsmath,amssymb,amsthm,mathtools}
\usepackage[margin=1in]{geometry}
\usepackage[expansion=false]{microtype}
\usepackage[authoryear,round]{natbib}
\usepackage[dvipsnames]{xcolor}
\usepackage[colorlinks=true,linkcolor=MidnightBlue,citecolor=MidnightBlue,
            urlcolor=MidnightBlue,hyperfootnotes=false]{hyperref}
\usepackage{setspace,booktabs}
\usepackage{pgfplots}
\usepgfplotslibrary{groupplots}
\pgfplotsset{compat=1.18}
\hypersetup{
  pdftitle={The Price of a Familiar Perspective},
  pdfauthor={Georgy Lukyanov and Ilya Pototskiy},
  pdfsubject={Source learning, renewal pricing, and access to historical reports},
  pdfkeywords={information markets, source familiarity, archive access, feedback, Bayesian learning}
}

\newtheorem{lemma}{Lemma}
\newtheorem{proposition}{Proposition}
\newtheorem{corollary}{Corollary}
\newtheorem{remark}{Remark}

\newcommand{\E}{\mathbb{E}}

\newcommand{\N}{\mathcal{N}}

\begin{document}

\setstretch{1.1}

\title{The Price of a Familiar Perspective}

\author{Georgy Lukyanov\thanks{Toulouse School of Economics, 1 Esplanade de l'Universit\'e, 31080 Toulouse, France. Corresponding author: \href{mailto:georgy.lukyanov@tse-fr.eu}{georgy.lukyanov@tse-fr.eu}.}
\and Ilya Pototskiy\thanks{International College of Economics and Finance, HSE University, 11 Pokrovsky Boulevard, 109028 Moscow, Russia.}}
\date{7 October 2026}

\maketitle

\begin{abstract}
Access to past reports changes how consumers learn about competing information providers. A subscriber who has learned to interpret a familiar source may find an equally competent rival less useful. We study how opening the rival's archive changes renewal prices and the effect of outcome feedback. Better feedback strengthens the familiar source's advantage when rival reports are inaccessible, but can weaken it once some rival records are available. We derive the exact threshold in a Gaussian learning model with differentiated price competition. Opening further records narrows the renewal premium and raises consumer surplus. When the familiar source initially serves a sufficiently large share, however, partial opening can reduce aggregate forecasting quality and total surplus by drawing consumers toward a rival that remains less informative. Complete access improves both outcomes relative to every incomplete archive. These results concern a covered interior renewal market with inherited customer histories, before implementation costs. The results distinguish the quality of feedback from access to the records with which consumers combine it.
\end{abstract}


\bigskip

\section{Introduction}

An investor who has followed the same analyst for several years knows how to read her recommendations. Optimism in one report and caution in another need not be taken at face value, because the investor has learned how the analyst's prior views enter her judgments. A new analyst may be equally competent and still be harder to use. Making the new analyst's past reports available seems a natural way to help the investor compare the two. We ask how such access changes competition between information providers, and whether the resulting gains to consumers also improve their forecasts.

Consider a subscriber to a paid forecasting service. Her account contains reports she has already purchased, and subsequent outcomes help her interpret them. Before renewing, she may also be able to inspect some of a rival's historical forecasts. Better evidence about past outcomes and wider access to past reports do different things. Outcome feedback makes an available record more informative; archive access determines which source that record lets her learn about. Understanding a source can make its next report more useful even when the source itself does not become more accurate.\footnote{A known perspective can be subtracted from a report. In our benchmark, uncertainty about perspective reduces report value; the magnitude of a perspective that is already known does not. This separates understanding a source from agreement with it or confidence in its neutrality.}

This distinction yields two results. First, better feedback need not erode the familiar source's price premium. If consumers can inspect only reports they previously purchased, feedback reinforces its advantage. Access to some rival reports can reverse the effect: consumers may gain more from learning about the source they understand less well, even though they have fewer records from it. We characterize the threshold at which this happens. Opening additional rival records always narrows the premium in the benchmark.

Second, lower prices and better information about the rival need not improve aggregate forecasting quality or total surplus. A consumer can benefit from switching while choosing a report that remains less informative. If the familiar source initially serves a sufficiently large share of the cohort, a partial archive opening can induce enough switching to outweigh the direct informational gain. Complete access nevertheless improves forecasting quality, consumer surplus and total surplus relative to every incomplete archive. It removes the information gap between the sources. These comparisons hold within a covered interior renewal market and before implementation costs; they do not imply that every increment of access is harmful or that complete opening is optimal on net.

The model combines Gaussian source learning with differentiated price competition. A source's report combines a noisy signal of a new state with its stable perspective according to a known rule. Past reports teach consumers how to subtract that perspective and recover more information from the next report. We hold the stock of historical records fixed: consumers have $n$ reports from the familiar source and can consult $m\leq n$ reports from its rival. Feedback has the same precision for each accessible record. Prices are chosen after this evidence is processed. Customer histories, expertise and the reporting rule are given, and pricing is terminal or myopic. The analysis therefore concerns what firms can charge for an established informational relationship and how access changes the renewal decision.

Our starting point is \citet{SethiYildiz2016}. Their equations~(5)--(7) provide the Gaussian filter we use, and their Section~8 establishes how delayed outcome revelation reinforces attachment to previously observed sources. \citet{SaghafianTomlinBiller2022} study the choice between familiar and accurate Gaussian sources with unknown inference models and imperfect outcome revelation; \citet{SethiYildiz2026} examine the broader consequences of differences in perspective. We introduce access to rival records into a priced renewal market. The contribution is the feedback reversal under partial access, together with the distinction between the allocation effects of partial and complete opening. Appendix~\ref{app:overlap} shows that these qualitative effects also occur when the two sources' reports concern overlapping historical dates, although the exact feedback threshold changes.

The paper also relates to learning and customer histories in competition. \citet{Bagwell1990} studies informational differentiation as an entry barrier. \citet{VillasBoas2004,VillasBoas2006} examine learning, loyalty and competition with experience goods, and \citet{DeNijsRhodes2013} study behavior-based pricing in that setting. Correlated product values in \citet{ChenStantonThomas2024} allow experience with one product to inform consumers about its rivals. Here perspectives are initially independent, so learning about a rival requires its reports. \citet{Campbell2015} connects subscription pricing and open access to a freely available understanding of an original source's contents. Our archive instead helps a consumer learn a stable reporting transformation and interpret the next report about a new state.

The tension between consumer gains and efficient allocation has precedents as well. Competitive poaching can induce inefficient switching \citep{FudenbergTirole2000}; information about product values can make consumer gains from competition conflict with efficient matching \citep{ArmstrongZhou2022,BergemannBrooksMorris2025}. In our model, evidence changes the decision value of the reports themselves. We distinguish the resulting prediction loss from the change in surplus and derive exact market-share conditions for losses after a finite archive opening. These conditions explain how partial opening can be harmful even though complete opening is beneficial.

The same learning technology appears in \citet{LukyanovOgorodnikov2026}, who study a monopolist's dynamic choice of evidence precision, including the role of public calibration. We hold expertise fixed and study competing sources, access to their records and renewal prices. Paid news and financial commentary are applications, but the mechanism does not require a preference for congenial news, strategic slant or reputation for accuracy, as in other media models \citep{MullainathanShleifer2005,GentzkowShapiro2006}. It also differs from verification and selective disclosure by competing outlets in \citet{HsuEtAl2022}.\footnote{\citet{GentzkowWongZhang2025} study trust when source accuracy is uncertain and ideological priors affect the assessment of reports. In our model expertise is known, while perspective is unknown and can be subtracted once learned. Understanding a source consequently means learning how to decode its reports.}

Section~\ref{sec:learning} derives report value and familiarity, and Section~\ref{sec:market} introduces renewal pricing. Section~\ref{sec:institutions} compares feedback quality, archive access and public calibration. Section~\ref{sec:welfare} studies forecasting quality and surplus; Section~\ref{sec:implications} discusses interpretation and limitations. Proofs are in Appendix~\ref{app:proofs}.

\section{Reports and source learning}\label{sec:learning}

\subsection{Environment}

We distinguish a source's stable perspective from the changing state about which it reports. Time is discrete. In every period $t$, a new state $\theta_t$ is drawn independently from
\[
\theta_t\sim\N(0,1).
\]
An information source $i$ has a fixed perspective $\mu_i$. The source's own prior for the current state is $\N(\mu_i,1)$. From the consumer's perspective,
\[
\mu_i\mid\mathcal H\sim\N(m_i,\rho_i^{-1}),
\]
where $\mathcal H$ is the consumer's information before the current report. The posterior mean $m_i$ is her estimate of the perspective, and $\rho_i>0$ measures how well she understands it.\footnote{Consider a tennis correspondent who admires a particular player. He may overstate the player's chance of winning after noisy news about fitness, or too readily attribute a defeat to temporary injury. A regular reader learns how much optimism to subtract. A newcomer may be just as aware of the correspondent's expertise, yet lack this understanding of his perspective, $\mu_i$.}

Source $i$ observes
\[
x_{it}=\theta_t+\varepsilon_{it},
\qquad
\varepsilon_{it}\sim\N(0,\phi^{-1}),
\]
where $\phi>0$ is expertise. The source then reports her posterior mean,
\begin{equation}\label{eq:report}
y_{it}=\frac{\mu_i+\phi x_{it}}{1+\phi}.
\end{equation}
Initial perspectives, states, and signal errors are mutually independent, and errors are independent across sources and dates. Expertise and the reporting rule are public. The source knows her own perspective. We hold the reporting transformation fixed and allow providers to compete through prices. This isolates the effect of consumers learning how to interpret a source.\footnote{The reporting rule can also describe a committed editorial technology, such as an outlet's house style. If a source could change its slant in response to being decoded, its reporting incentives and consumers' learning would have to be determined jointly.}

After observing any reports purchased in period $t$, the consumer chooses an action $a_t$ and receives payoff
\[
-(a_t-\theta_t)^2.
\]
The optimal action is the posterior mean. We measure the gross value of a report by the ex ante reduction in the consumer's expected quadratic loss relative to receiving no report. The state and realized loss are not observed directly; Section~\ref{sec:institutions} introduces a noisy post-action signal of the state.

\subsection{The value of one report}

A consumer uses her estimate of the perspective to recover the information in a report. After subtracting the estimated perspective and rescaling, she obtains
\begin{equation}\label{eq:zsignal}
z_{it}\equiv\frac{(1+\phi)y_{it}-m_i}{\phi}
=\theta_t+\varepsilon_{it}+\frac{\mu_i-m_i}{\phi}.
\end{equation}
Conditional on $\mathcal H$, this is a signal of $\theta_t$ with independent noise variance
\[
\frac{1}{\phi}+\frac{1}{\phi^2\rho_i}.
\]
Its precision is therefore
\begin{equation}\label{eq:lambda}
\lambda(\rho_i,\phi)
=\frac{\phi^2\rho_i}{1+\phi\rho_i}.
\end{equation}

\begin{samepage}
\begin{lemma}\label{lem:value}
Given perspective precision $\rho>0$, the posterior variance of the state after one report is
\[
v(\rho,\phi)=\frac{1}{1+\lambda(\rho,\phi)}
=\frac{1+\phi\rho}{1+\phi(1+\phi)\rho}.
\]
The gross value of the report is
\begin{equation}\label{eq:b}
b(\rho,\phi)
=1-v(\rho,\phi)
=\frac{\phi^2\rho}{1+\phi(1+\phi)\rho}.
\end{equation}
It is increasing and strictly concave in $\rho$, and increasing in $\phi$.
\end{lemma}
\end{samepage}

Expertise reduces private-signal noise; understanding reduces the error in subtracting perspective. The concavity of \eqref{eq:b} means that further understanding is worth less when the perspective is already well known. If it were known exactly, the report would reveal the source's signal, worth $\phi/(1+\phi)$. Familiarity can approach this value but cannot improve the underlying signal.

\subsection{Learning a perspective}

The same report also teaches the consumer something about the source. Multiplying \eqref{eq:report} by $1+\phi$ gives
\begin{equation}\label{eq:musignal}
(1+\phi)y_{it}
=\mu_i+\phi\theta_t+\phi\varepsilon_{it}.
\end{equation}
Suppose first that the realized state is never separately revealed. Conditional on $\mu_i$, the last two terms in \eqref{eq:musignal} have variance $\phi^2+\phi$. One exposure therefore adds perspective precision
\begin{equation}\label{eq:kappa0}
\kappa_0(\phi)=\frac{1}{\phi(1+\phi)}.
\end{equation}
After $n$ reports from the same identified source, the consumer's precision about its perspective is $\rho+n\kappa_0(\phi)$. Learning is possible without observing any realized state because the distribution of states is known and stationary. If that distribution were unknown or drifting, a persistent change in reports could instead reflect a change in the world. We set this identification problem aside.

The value of familiarity after $n$ exposures is the difference between the value of a report from the familiar source and that of an otherwise identical unfamiliar source:
\begin{equation}\label{eq:delta_def}
\Delta_n(\rho,\phi)
\equiv b\bigl(\rho+n\kappa_0(\phi),\phi\bigr)-b(\rho,\phi).
\end{equation}

\begin{samepage}
\begin{proposition}\label{prop:familiarity}
For every $n\geq1$, $\rho>0$, and $\phi>0$,
\begin{equation}\label{eq:delta_closed}
\Delta_n(\rho,\phi)
=\frac{n\phi}
{(1+\phi)[1+\rho\phi(1+\phi)]
[n+1+\rho\phi(1+\phi)]}.
\end{equation}
The familiarity advantage is positive, increases at a diminishing rate with $n$, and decreases with $\rho$. As a function of expertise, it has a unique interior maximum.
\end{proposition}
\end{samepage}

Repeated exposure adds the same precision each time, but its value diminishes as the source becomes better understood. Familiarity matters most at intermediate expertise: poor sources provide little information worth recovering, while very expert sources are useful even to a newcomer. The maximizing expertise and the diffuse-prior boundary are characterized in Appendix~\ref{app:expertise}.

\section{Renewal pricing}\label{sec:market}

Consider a cohort that has previously obtained $n$ reports from source $F$ and none from source $U$. Cohort membership is observable to both sources, offers can be conditioned on it, and there is no arbitrage across cohorts. A free trial assigned independently of horizontal preferences is one interpretation. The pricing game is terminal; equivalently, firms are myopic about any source learning their current sales create. Both sources have expertise $\phi$, marginal cost $c\geq0$, and the same initial perspective precision $\rho$. Prices cannot fall below $c$.

Historical evidence is processed before prices are chosen. The current state, signals, and reports are realized afterward. We study a non-signalling price game: commercial pricing units observe the common evidence available to the cohort but have no private information about the perspectives. The reporting routine can use its own perspective, as in \eqref{eq:report}, but cannot communicate it separately before purchase. Thus a price conveys no additional information about report value.\footnote{This separation is substantive. Verifiable disclosure of the perspective could remove calibration uncertainty; a privately informed seller's price could itself convey information. Neither disclosure nor a price-signalling game is included in the benchmark.}

Let $b_F\geq b_U$ be the report values when prices are chosen, and write $\Delta=b_F-b_U$. With only the inherited history, $b_F=b(\rho+n\kappa_0,\phi)$ and $b_U=b(\rho,\phi)$, so the gap is $\Delta_n$ from Proposition~\ref{prop:familiarity}. Section~\ref{sec:institutions} changes these values by supplying public profiles, outcome feedback, or access to an archive. The pricing result below applies to all these cases. Realized posterior means may differ, but report values depend only on the corresponding precisions and are common within the cohort.

Consumers are indexed by $x\in[0,1]$ and distributed uniformly. Source $F$ is located at $0$ and source $U$ at $1$. A consumer's utility from purchasing one report is
\begin{align*}
u_F(x)&=\bar v+b_F-p_F-\tau x,\\
u_U(x)&=\bar v+b_U-p_U-\tau(1-x),
\end{align*}
where $\tau>0$ captures horizontal differences in fit, presentation, or topic coverage, in the manner of \citet{Hotelling1929}. Consumers buy at most one current report. This is a maintained restriction on current consumption, motivated by limited attention.\footnote{Consumers have very little time and a great deal of available information. Processing a second report can improve a decision and still not be worth the attention it requires. We do not derive single-homing as an equilibrium of the differentiated market studied here.} Each price buys access to one new report; for a paid outlet, this can be read as the price of a short renewal interval. The common component $\bar v$ is large enough for the candidate market to be covered, and sources set prices simultaneously.

\begin{samepage}
\begin{proposition}\label{prop:hotelling}
Suppose $0\leq\Delta<3\tau$ and
\[
\bar v+b_U-c-\frac{3\tau}{2}+\frac{\Delta}{2}>0.
\]
The unique covered interior price equilibrium is
\begin{equation}\label{eq:prices}
p_F^*=c+\tau+\frac{\Delta}{3},
\qquad
p_U^*=c+\tau-\frac{\Delta}{3}.
\end{equation}
The familiar source serves the share
\begin{equation}\label{eq:share}
x_F^*=\frac{1}{2}+\frac{\Delta}{6\tau}
\end{equation}
and charges the price premium
\[
p_F^*-p_U^*=\frac{2\Delta}{3}.
\]
\end{proposition}
\end{samepage}

For a positive gap, the familiar source charges more and serves a larger share of its returning cohort. Two thirds of the informational advantage appears in the price difference in this Hotelling specification; the remainder shifts the marginal consumer. The fraction depends on demand structure. These are cohort-specific comparisons. With mirror cohorts of legacy $F$ and legacy $U$ consumers, both firms can charge their own returning users more and retain a disproportionate share while aggregate firm shares remain symmetric.

\section{Feedback and access to historical reports}\label{sec:institutions}

\subsection{A common technology for learning from records}

The cohort enters the renewal stage with $n\geq1$ stored reports from $F$. An archive contains $n$ past reports from $U$, of which $m\in\{0,\ldots,n\}$ are accessible before renewal. These rival reports concern dates not represented in the stored $F$ history. Their states and signal errors are therefore independent of the cohort's historical evidence about $F$. Archive selection depends on the number of records, not their contents. This permits a transparent comparison of access while holding the available stock of records fixed.\footnote{Reports about shared historical states require joint filtering. Appendix~\ref{app:overlap} derives that filter and shows that the feedback reversal and the contrast between partial and complete opening survive. The exact threshold below is specific to independent dates and cannot be applied mechanically to overlapping records.}

For each accessible report, a truthful outcome record is supplied after the historical action and before renewal prices are set:
\[
d_t=\theta_t+\nu_t,
\qquad \nu_t\sim\N(0,\delta^{-1}).
\]
The errors are independent across dates and of all other primitives. The precision $\delta\geq0$ is the same for both sources. At $\delta=0$ there is no outcome information; as $\delta\to\infty$, the historical state becomes known. The consumer can consult an archived report for calibration even though it was not purchased when the historical decision was made. Access to these records, or an equivalent sufficient statistic, is costless to the consumer in the benchmark. Section~\ref{subsec:costs} discusses implementation costs.

Conditional on $d_t$, the variance of the historical state is $1/(1+\delta)$. The residual
\[
(1+\phi)y_{it}-\phi\E[\theta_t\mid d_t]
=\mu_i+\phi\bigl(\theta_t-\E[\theta_t\mid d_t]\bigr)+\phi\varepsilon_{it}
\]
has noise variance $\phi+\phi^2/(1+\delta)$. Each report--outcome pair therefore supplies precision
\begin{equation}\label{eq:kappadelta}
\kappa(\delta,\phi)
=\frac{1+\delta}{\phi(1+\delta+\phi)},
\qquad
\kappa_\delta=\frac{1}{(1+\delta+\phi)^2}>0.
\end{equation}
At zero feedback this is $\kappa_0$; under exact feedback it approaches $1/\phi$. This is the precision increment in \citet[equation~(7)]{SethiYildiz2016}, with feedback precision in place of the observer's private-signal precision.

We also allow an independent public profile $s_i^\chi=\mu_i+\eta_i$ for each source, where $\eta_i\sim\N(0,\chi^{-1})$. The precision $\chi\geq0$ and access to the profiles are symmetric; $\chi=0$ denotes their absence. The profile errors are independent of each other and of all other primitive shocks. Profiles and historical records are conditionally independent given the perspectives. They estimate perspective, rather than rate neutrality or honesty.

Before renewal, the perspective precisions and report values are
\begin{equation}\label{eq:archive_precisions}
r_F=\rho+\chi+n\kappa(\delta,\phi),\qquad
r_U=\rho+\chi+m\kappa(\delta,\phi),\qquad
b_i=b(r_i,\phi).
\end{equation}
The $F$ precision in \eqref{eq:archive_precisions} replaces the precision from its original $n$ reports. The reports are processed again with their outcomes; they are not counted a second time. More precisely, feedback adds $n(\kappa-\kappa_0)$ to the inherited precision $\rho+n\kappa_0$.

All evidence is processed before firms choose prices, and accessible record counts are common knowledge. The Gaussian posterior variances do not depend on realized record values. Consequently the pricing game still depends on the information only through
\begin{equation}\label{eq:archive_gap}
\Delta_{n,m}(\chi,\delta)=b(r_F,\phi)-b(r_U,\phi).
\end{equation}
The label $U$ continues to identify the rival, even when an archive makes it partially familiar. For $m<n$, $F$ remains better understood. At $m=n$, report values are equal.

\subsection{When better feedback strengthens familiarity}

Let
\begin{equation}\label{eq:access_variables}
q=\phi(1+\phi),\qquad A=1+q(\rho+\chi),\qquad
\zeta=q\kappa(\delta,\phi)
=\frac{(1+\phi)(1+\delta)}{1+\delta+\phi}.
\end{equation}
Here $\zeta$ increases from $1$ toward $1+\phi$. It measures the information supplied by a record relative to a report without outcome feedback.

\begin{samepage}
\begin{proposition}\label{prop:access_feedback}
For $0\leq m<n$,
\begin{equation}\label{eq:access_closed}
\Delta_{n,m}
=\frac{\phi}{1+\phi}
\frac{(n-m)\zeta}{(A+n\zeta)(A+m\zeta)}>0.
\end{equation}
If $m=0$, better feedback strictly increases $\Delta_{n,0}$. If $0<m<n$, its marginal effect has the sign
\begin{equation}\label{eq:feedback_sign}
\operatorname{sgn}\!\left(\frac{\partial\Delta_{n,m}}{\partial\delta}\right)
=\operatorname{sgn}(A^2-nm\zeta^2).
\end{equation}
In particular, writing $H=A/\sqrt{nm}$:
\begin{enumerate}
\item If $H\leq1$, the advantage decreases with feedback; the derivative at zero is zero only when $H=1$.
\item If $H\geq1+\phi$, the advantage strictly increases at every finite feedback precision.
\item If $1<H<1+\phi$, it increases and then decreases, with a unique maximum at
\begin{equation}\label{eq:feedback_peak}
\delta^*=\frac{(1+\phi)(H-1)}{1+\phi-H}.
\end{equation}
\end{enumerate}
At $m=n$, the advantage is identically zero.
\end{proposition}
\end{samepage}

Better feedback helps consumers interpret every accessible report. Since they have more reports from $F$, this gives the familiar source an advantage. At the same time, they already understand $F$ better, so an additional unit of precision is less valuable there. With $m=0$, the rival has no report that the consumer can pair with the outcome, and only the familiar source benefits. Once rival records become available, the larger gain from learning about a less understood source can outweigh the difference in record counts.

Within the covered interior market, the price gap, $F$'s share, and total industry profit move with $\Delta_{n,m}$. Thus \eqref{eq:feedback_sign} also determines their response to feedback. A decreasing advantage does not mean that the rival becomes more informative: $F$ has the larger report value whenever $m<n$. For example, at $\phi=1$, $\rho=0.1$, $\chi=0$, and $n=5$, feedback strengthens familiarity when $m=0$ but weakens it at every finite precision when even one rival report is accessible. With $\phi=1$, $\rho=0.5$, $\chi=0$, $n=2$, and $m=1$, the interior maximum occurs at $\delta^*=\sqrt{2}$.

\subsection{Opening an archive and supplying public profiles}

Archive access and feedback quality are different instruments. Opening a further rival record leaves $b_F$ unchanged and raises $b_U$. It therefore narrows the familiarity advantage at every feedback precision. A symmetric public profile improves both report values, but helps the less understood source more.

\begin{samepage}
\begin{corollary}\label{cor:access_calibration}
Holding feedback precision fixed, increasing $m$ strictly reduces $\Delta_{n,m}$ until it reaches zero at $m=n$. Holding record counts fixed with $m<n$, increasing $\chi$ strictly reduces $\Delta_{n,m}$. In the covered interior equilibrium, either change lowers $F$'s price, share, and profit; raises $U$'s price, share, and profit; and lowers total industry profit.
\end{corollary}
\end{samepage}

Outcome feedback and public profiles thus differ in the evidence they make usable. Without rival records, feedback improves understanding only of the source the consumer has used. Public profiles improve understanding of both sources and help the less familiar one more. Proposition~\ref{prop:access_feedback} shows how opening an archive changes the first comparison: feedback can now be applied to both sources, so its competitive effect depends on the extent of access.

\begin{samepage}
\begin{remark}\label{rem:architecture}
The mechanism behind the sign comparison extends beyond Gaussian learning. For $r>0$, $k>0$, $n>m\geq0$, and a differentiable, increasing and strictly concave report-value function $B$, the advantage $B(r+nk)-B(r+mk)$ falls after an equal increment to $r$. Its response to better records, represented by $k$, is
\[
nB'(r+nk)-mB'(r+mk).
\]
At $m=0$ it is positive. For $m>0$, concavity alone does not determine its sign. The Gaussian specification gives the closed-form threshold in \eqref{eq:feedback_sign}; it is not needed for the distinction between record quality and record access.
\end{remark}
\end{samepage}
\section{Forecasting quality and renewal surplus}\label{sec:welfare}

A smaller renewal premium need not mean better forecasts or higher total surplus. To see why, we distinguish the reduction in aggregate prediction loss, consumer surplus after prices and horizontal mismatch, and total surplus with prices treated as transfers. Each measure concerns the renewal period, conditional on the inherited history. The distinction matters when the familiar source serves a large share of the cohort. For independent historical dates, a finite archive opening lowers forecasting benefit exactly when the average of its initial and final shares exceeds three quarters; the corresponding cutoff for total surplus is four fifths. Below these cutoffs, the direct benefit of access is not outweighed by the change in source choice. We derive these conditions below.

\subsection{Prices and the allocation of consumers}

Let $\Delta=b_F-b_U\geq0$. If a share $x$ is assigned to $F$, gross renewal-period surplus is
\begin{equation}\label{eq:welfare_x}
W(x)=\bar v+b_U-c+\Delta x
-\frac{\tau}{2}\bigl[x^2+(1-x)^2\bigr].
\end{equation}
The last term is aggregate mismatch cost. The planner takes the two report values as given and chooses who receives each report.

\begin{samepage}
\begin{proposition}\label{prop:allocation}
Under the conditions of Proposition~\ref{prop:hotelling}, the surplus-maximizing share is
\begin{equation}\label{eq:efficient_share}
x_F^S=\min\left\{1,\frac12+\frac{\Delta}{2\tau}\right\}.
\end{equation}
For $0<\Delta<3\tau$, equilibrium assigns strictly fewer consumers to $F$ than the planner does. The deadweight loss is
\begin{equation}\label{eq:dwl}
DWL=
\begin{cases}
\displaystyle \frac{\Delta^2}{9\tau}, & 0\leq\Delta<\tau,\\[5pt]
\displaystyle \frac{(3\tau-\Delta)(5\Delta-3\tau)}{36\tau},
& \tau\leq\Delta<3\tau.
\end{cases}
\end{equation}
\end{proposition}
\end{samepage}

The familiar source captures part of its informational advantage through its price. Some consumers consequently buy the rival's report even though the gain from a better forecast would justify their additional mismatch cost at $F$. Better information changes report values, but it can also change this allocation distortion.

\subsection{Three measures of the gains from information}

At the equilibrium share $x_F^*=1/2+\Delta/(6\tau)$, aggregate forecasting benefit is
\begin{equation}\label{eq:forecasting}
Q^E=(1-x_F^*)b_U+x_F^*b_F
=b_U+\frac{\Delta}{2}+\frac{\Delta^2}{6\tau}.
\end{equation}
Since the prior variance of the state is one, aggregate expected quadratic prediction loss is $1-Q^E$. Consumer surplus and gross total surplus are respectively
\begin{align}
CS^E&=\bar v+b_U-c-\frac{5\tau}{4}
+\frac{\Delta}{2}+\frac{\Delta^2}{36\tau},\label{eq:cs}\\
W^E&=\bar v+b_U-c-\frac{\tau}{4}
+\frac{\Delta}{2}+\frac{5\Delta^2}{36\tau}.\label{eq:welfare_equilibrium}
\end{align}
Their difference is total industry profit,
\begin{equation}\label{eq:industryprofit}
W^E-CS^E=\tau+\frac{\Delta^2}{9\tau}.
\end{equation}

\begin{samepage}
\begin{proposition}\label{prop:performance}
Consider a differentiable parameter change $s$ with $b_{F,s},b_{U,s}\geq0$ and $b_{F,s}+b_{U,s}>0$, preserving the covered interior equilibrium. Its effects are
\begin{equation}\label{eq:performance_weights}
\frac{dJ^E}{ds}=(1-\omega_J)b_{U,s}+\omega_J b_{F,s},
\qquad J\in\{CS,Q,W\},
\end{equation}
where
\[
\omega_{CS}=\frac12+\frac{\Delta}{18\tau},\qquad
\omega_Q=\frac12+\frac{\Delta}{3\tau},\qquad
\omega_W=\frac12+\frac{5\Delta}{18\tau}.
\]
Consumer surplus strictly increases. Forecasting benefit and gross total surplus can decrease only when the change narrows the advantage. A decrease in $Q^E$ additionally requires $x_F^*>3/4$; a decrease in $W^E$ requires $x_F^*>4/5$.
\end{proposition}
\end{samepage}

The consumer-surplus weights form a convex combination because $1/2\leq\omega_{CS}<2/3$: improving either report benefits consumers. The weights for forecasting and total surplus can exceed one because they also reflect the change in source choice. A consumer who switches may gain through a lower price or a better horizontal fit while receiving a report with a larger prediction error. If enough consumers switch, this loss can outweigh the direct improvement in the rival's report.

Better feedback raises both report values when $m>0$, and only $F$'s value when $m=0$. It always raises consumer surplus. In the closed-archive case it also raises $Q^E$ and $W^E$. With partial access, the feedback threshold in Proposition~\ref{prop:access_feedback} determines whether there is a potentially adverse reallocation effect; a narrowing advantage is necessary, but not sufficient, for a forecasting or welfare loss.

For a public profile, define
\begin{equation}\label{eq:t_ratio}
t_\chi=\frac{b_{F,\chi}}{b_{U,\chi}}
=\left(\frac{A+m\zeta}{A+n\zeta}\right)^2.
\end{equation}
For $m<n$, this ratio lies strictly between zero and one. The exact conditions are
\begin{align}
Q^E_\chi<0
&\quad\Longleftrightarrow\quad
t_\chi<\frac{2\Delta-3\tau}{2\Delta+3\tau},\label{eq:calibration_forecast}\\
W^E_\chi<0
&\quad\Longleftrightarrow\quad
t_\chi<\frac{5\Delta-9\tau}{5\Delta+9\tau}.\label{eq:calibration_welfare}
\end{align}
Equality gives a zero marginal effect. A negative right-hand side cannot exceed $t_\chi$, so these expressions include the regions in which a loss is impossible. Public profiles always improve each source's report at a fixed choice, but their aggregate prediction effect can be negative. For instance, $\phi=1$, $\rho=0.1$, $n=5$, $m=\chi=\delta=0$, and $\tau=0.16$ give $Q^E_\chi\simeq-0.1077$, $CS^E_\chi\simeq0.2822$, and $W^E_\chi\simeq-0.0297$. Taking $\bar v=2$ and $c=0$ satisfies coverage. The derivatives at zero are right derivatives.

\subsection{Partial and complete archive opening}

Opening an archive is a discrete change in record access. It need not be approximated by a marginal change in precision. Hold $n$, $\chi$, and $\delta$ fixed, and increase accessible rival records from $m_0$ to $m_1$, where $0\leq m_0<m_1\leq n$. Let $\Delta_0$ and $\Delta_1$ be the associated advantages and put $d=\Delta_0-\Delta_1>0$. This is also the increase in $b_U$; $b_F$ is unchanged.

\begin{samepage}
\begin{proposition}\label{prop:archive_welfare}
Suppose the initial equilibrium is covered and interior. Every increase in archive access preserves coverage and an interior market share, including the symmetric outcome at $m=n$. The finite changes are
\begin{align}
CS^E_1-CS^E_0
&=d\left[\frac12-\frac{\Delta_0+\Delta_1}{36\tau}\right]>0,\label{eq:archive_cs}\\
Q^E_1-Q^E_0
&=d\left[\frac12-\frac{\Delta_0+\Delta_1}{6\tau}\right],\label{eq:archive_q}\\
W^E_1-W^E_0
&=d\left[\frac12-\frac{5(\Delta_0+\Delta_1)}{36\tau}\right].\label{eq:archive_w}
\end{align}
Thus forecasting benefit falls if and only if the average of the initial and final familiar-source shares exceeds $3/4$. Gross total surplus falls if and only if their average exceeds $4/5$. Complete opening, $m_1=n$, strictly raises all three outcomes relative to every incomplete archive.
\end{proposition}
\end{samepage}

With independent historical dates, opening rival records improves $b_U$ while leaving $b_F$ fixed. The share conditions describe when the resulting movement of consumers is costly. A partial opening makes the rival easier to understand, but its report is still less informative. Drawing consumers toward it can then reduce aggregate forecasting benefit and gross surplus. Complete opening removes this informational disadvantage. The reports become equally valuable, their prices are equal, and consumers choose the source that best suits their horizontal preferences.

Figure~\ref{fig:archive} illustrates all six access levels for a fixed stock of five reports per source. Opening the first rival record lowers forecasting benefit and gross surplus, even though consumer surplus rises. Further opening reverses these losses; complete access yields the highest value of each outcome. Initially, the familiar source serves about $95\%$ of the cohort. The average of its shares before and after the first opening is about $82\%$, above both loss cutoffs. The example therefore illustrates the conditions in Proposition~\ref{prop:archive_welfare}, rather than a general presumption against partial access. Feedback quality is fixed throughout. Appendix~\ref{app:overlap} obtains the same qualitative comparison when reports concern overlapping historical dates, although opening the archive then improves both report values.

\begin{figure}[htbp]
\centering
\begin{tikzpicture}
\begin{groupplot}[
group style={group size=3 by 1,horizontal sep=1.3cm},
width=0.30\textwidth,height=5.1cm,
xmin=-0.15,xmax=5.15,xtick={0,1,2,3,4,5},
xlabel={Rival records $m$},
tick label style={font=\footnotesize},
label style={font=\small},title style={font=\small},
scaled y ticks=false,y tick label style={/pgf/number format/fixed,/pgf/number format/precision=3},
axis line style={black!60},tick style={black!60},
grid=major,grid style={black!10},
]
\nextgroupplot[title={Consumer surplus},ymin=-0.008,ymax=0.155,ytick={0,0.05,0.1,0.15}]
\addplot[black!55,dashed,domain=0:5] {0};
\addplot[black,thick,mark=*,mark size=2pt] coordinates {(0,0.000000000000) (1,0.074383456593) (2,0.106387364347) (3,0.124046034184) (4,0.135218647419) (5,0.142919541116)};
\nextgroupplot[title={Forecasting benefit},ymin=-0.032,ymax=0.023,ytick={-0.03,-0.015,0,0.015}]
\addplot[black!55,dashed,domain=0:5] {0};
\addplot[black,thick,mark=*,mark size=2pt] coordinates {(0,0.000000000000) (1,-0.027184108929) (2,-0.012717480585) (3,0.000228586058) (4,0.010029833231) (5,0.017463483254)};
\nextgroupplot[title={Gross total surplus},ymin=-0.011,ymax=0.047,ytick={-0.01,0,0.02,0.04}]
\addplot[black!55,dashed,domain=0:5] {0};
\addplot[black,thick,mark=*,mark size=2pt] coordinates {(0,0.000000000000) (1,-0.006870595825) (2,0.011103488401) (3,0.024992075683) (4,0.035067596068) (5,0.042554694827)};
\end{groupplot}
\end{tikzpicture}
\caption{Changes from a closed archive}\label{fig:archive}\label{tab:archive}
\begin{minipage}{0.97\textwidth}
\footnotesize
Notes: Each panel reports the change from $m=0$, in payoff units; vertical scales differ. The six points are the feasible record counts. Connecting lines guide the eye. Parameters are $\phi=1$, $\rho=0.1$, $n=5$, $\chi=\delta=0$, $\tau=0.125$, $\bar v=2$, and $c=0$. No outcomes are revealed at $\delta=0$, but archived reports still inform consumers about perspective. Coverage and interiority hold at every point. Higher forecasting benefit means lower expected prediction loss. Surplus excludes implementation costs.
\end{minipage}
\end{figure}

\subsection{Implementation costs}\label{subsec:costs}

Let $K(\chi,\delta,m;n)$ be the allocated or amortized resource cost of providing profiles, outcome feedback, and access to records, measured in renewal-period units. Net surplus is
\begin{equation}\label{eq:net_welfare}
\mathcal W=W^E-K(\chi,\delta,m;n).
\end{equation}
An archive opening is worthwhile on net only if its gross benefit exceeds its incremental resource cost. Starting from a covered interior equilibrium, complete access maximizes gross surplus among the access levels considered, but it need not maximize net surplus. The same distinction applies to public profiles and better outcome feedback: their marginal gross benefits must cover their marginal costs. These comparisons take the provision of information as given. A cost technology, financing mechanism, and model of private provision would be needed to determine which institution is supplied or to evaluate policy over the lifetime of a customer relationship.
\section{Interpretation and limitations}\label{sec:implications}

The model applies most directly to paid forecasting, financial commentary, newsletters, and other information services with identifiable repeat customers. In each case, a stable perspective can reflect an analyst's prior beliefs or a service's reporting convention. Learning that perspective lets a consumer extract more information from the next report, even though the source's expertise is unchanged.

The access comparison distinguishes a private subscriber history from an archive that can be consulted before purchase. A resolved forecast stored only in the subscriber's account informs the consumer about a source she has used. Making a rival's historical forecasts available permits outcome evidence to inform switching decisions as well. A public profile provides another way to learn about a source without having consumed its reports. These examples concern evidence about a source's reporting transformation. A generic credibility label need not estimate perspective or have the same effect.\footnote{A profile in this model resembles a description of how to read a source, rather than a seal certifying that the source is correct. A precise description of a pronounced perspective can be useful even when readers disagree with it.}

The mechanism suggests several ways to distinguish interpretive familiarity from other sources of loyalty. Holding expertise and horizontal fit fixed, exclusive past exposure raises the familiarity premium at a diminishing rate, whereas public information about perspective narrows it. Better outcome feedback strengthens the premium when rival reports are inaccessible, but can weaken it once consumers can consult those reports. Varying record access separately from feedback quality would therefore help identify the mechanism. Retention alone would not distinguish it from habit or a preference for congenial content.

Three restrictions are central to interpretation. First, the inherited cohort and its information are exogenous. Prices are terminal or myopic, so the model does not account for investment in acquiring customers or strategic provision of trial reports. Second, the exact feedback threshold and finite-opening formulas use records about independent dates, selected without regard to content. Appendix~\ref{app:overlap} treats overlapping dates: the qualitative effects survive, but the threshold does not carry over. Strategic disclosure and price signalling are outside the maintained information structure. Third, consumers purchase at most one current report. The analysis does not solve the interaction between multihoming, differentiated pricing, and future source learning. These restrictions define a conditional renewal problem in which the consequences of access can be stated precisely.

\section{Conclusion}\label{sec:conclusion}

Access to historical reports changes the competitive value of familiarity. Better outcome feedback reinforces a source's renewal premium when consumers cannot consult its rival's records. Once some rival reports are available, the same feedback can narrow the premium by helping consumers learn about the source they understand less well. The distinction is between improving evidence and making that evidence usable for a different choice.

Opening further rival records narrows the premium and raises consumer surplus in the covered interior market. Its effect on forecasting quality and total surplus depends on how consumers change sources. When the familiar source has a sufficiently large advantage, partial opening can draw consumers toward a rival that remains less informative and reduce both outcomes. Complete access removes this information gap and improves all three outcomes relative to every incomplete archive. These are gross gains at the renewal stage: implementation costs, customer acquisition and incentives to supply access remain separate questions.

For forecasting services and other information providers, a record of past reports is therefore more than evidence of accuracy. It also teaches a consumer how to read the next report. Whether outcome disclosure strengthens or weakens an established relationship depends on whose records the consumer can inspect.

\appendix

\section{Proofs}\label{app:proofs}

\subsection{Proof of Lemma~\ref{lem:value}}

From \eqref{eq:zsignal}, the report is equivalent to a signal of $\theta_t$ with noise variance
\[
\sigma_z^2=\frac{1}{\phi}+\frac{1}{\phi^2\rho}.
\]
The associated precision is $\lambda=1/\sigma_z^2=\phi^2\rho/(1+\phi\rho)$. Since the prior precision of $\theta_t$ is one, posterior variance is $(1+\lambda)^{-1}$, which gives the expression for $v$ and hence \eqref{eq:b}. Writing $q=\phi(1+\phi)$,
\[
\frac{\partial b}{\partial\rho}
=\frac{\phi^2}{(1+q\rho)^2}>0,
\qquad
\frac{\partial^2 b}{\partial\rho^2}
=-\frac{2q\phi^2}{(1+q\rho)^3}<0.
\]
Moreover,
\[
\frac{\partial b}{\partial\phi}
=\frac{\rho\phi(2+\rho\phi)}{[1+\rho\phi(1+\phi)]^2}>0.
\]
\qed

\subsection{Proof of Proposition~\ref{prop:familiarity}}\label{app:expertise}

The closed form follows from a change of variables that is worth isolating, because it is what makes the whole comparative-static analysis tractable. Let
\[
q=\phi(1+\phi),
\qquad
A=1+q\rho.
\]
Equation \eqref{eq:b} can be written as
\[
b(\rho,\phi)
=\frac{\phi}{1+\phi}\left(1-\frac{1}{A}\right).
\]
Because $q\kappa_0=1$, replacing $\rho$ by $\rho+n\kappa_0$ replaces $A$ by $A+n$: one exposure advances the state variable by exactly one unit, whatever the level of expertise. Hence
\[
\Delta_n
=\frac{\phi}{1+\phi}
\left(\frac{1}{A}-\frac{1}{A+n}\right)
=\frac{n\phi}{(1+\phi)A(A+n)},
\]
which is \eqref{eq:delta_closed}. Extending $n$ to the positive real line gives
\[
\frac{\partial\Delta_n}{\partial n}
=\frac{\phi}{(1+\phi)(A+n)^2}>0,
\qquad
\frac{\partial^2\Delta_n}{\partial n^2}
=-\frac{2\phi}{(1+\phi)(A+n)^3}<0.
\]
The derivative with respect to $\rho$ is
\[
\frac{\partial\Delta_n}{\partial\rho}
=b_\rho(\rho+n\kappa_0,\phi)-b_\rho(\rho,\phi)<0
\]
by strict concavity of $b$ in its first argument.

The maximizing expertise $\phi_n^*$ is the unique positive solution to
\begin{equation}\label{eq:peak}
n+1
=2(n+2)\rho\phi^2(1+\phi)
+\rho^2\phi^2(1+\phi)^2(1+4\phi).
\end{equation}
To establish this characterization, logarithmic differentiation gives
\[
\frac{\partial\log\Delta_n}{\partial\phi}
=\frac{D_n(\phi)}
{\phi(1+\phi)A(A+n)},
\]
where
\[
D_n(\phi)
=n+1-2(n+2)\rho\phi^2(1+\phi)
-\rho^2\phi^2(1+\phi)^2(1+4\phi).
\]
For $\rho>0$, $D_n$ is strictly decreasing on the positive real line, starts at $n+1$, and tends to minus infinity. It therefore has one positive zero, characterized by \eqref{eq:peak}. Finally, \eqref{eq:delta_closed} tends to zero as $\phi$ tends to either zero or infinity, so the unique stationary point is the unique global maximum. \qed

The restriction $\rho>0$ is important. At the improper-prior boundary $\rho=0$, \eqref{eq:delta_closed} is understood as a limit and equals $n\phi/[(n+1)(1+\phi)]$, which increases monotonically in expertise; the hump disappears. Moreover, as a proper prior becomes increasingly diffuse, the maximizing expertise moves arbitrarily far to the right. The result is exact for every $\rho>0$, but over a bounded empirical range familiarity may still appear to rise with expertise when initial knowledge of perspective is very imprecise.

\subsection{Proof of Proposition~\ref{prop:hotelling}}

Let $\Delta=b_F-b_U$. The indifferent consumer is located at
\begin{equation}\label{eq:cutoff_app}
x_F=\frac{\tau+\Delta-p_F+p_U}{2\tau}.
\end{equation}
On the interior, profits are $(p_F-c)x_F$ and $(p_U-c)(1-x_F)$. The first-order conditions are
\[
2(p_F-c)=\tau+\Delta+(p_U-c),
\qquad
2(p_U-c)=\tau-\Delta+(p_F-c).
\]
Solving gives \eqref{eq:prices}; substitution into \eqref{eq:cutoff_app} gives \eqref{eq:share}, and equilibrium profits are
\begin{equation}\label{eq:profits}
\pi_F^*=\frac{(3\tau+\Delta)^2}{18\tau},
\qquad
\pi_U^*=\frac{(3\tau-\Delta)^2}{18\tau}.
\end{equation}
Under $\Delta<3\tau$, both shares and markups are positive. To check global deviations, clip \eqref{eq:cutoff_app} to $[0,1]$. Interior profit is strictly concave. Given the candidate $p_U$, the most profitable price at which $F$ captures the entire market yields profit $2\Delta/3$, while its interior profit exceeds this by $(3\tau-\Delta)^2/(18\tau)$. Source $U$ could capture the entire market only at a price no greater than $c$, yielding no positive profit; losing the whole market yields zero. The stated inequality makes the marginal consumer's utility strictly positive at the candidate. An outside option can only reduce demand away from the clipped covered-market demand and therefore creates no profitable additional deviation. \qed

\subsection{Proof of Proposition~\ref{prop:access_feedback}}

Conditioning the report on its outcome gives the variance and precision in \eqref{eq:kappadelta}. Historical dates are distinct, so these signals are conditionally independent across the two source histories. A stored report with a newly observed outcome is one joint observation, whose total precision contribution is $\kappa$, rather than $\kappa_0+\kappa$. Independent public profiles add $\chi$ to each perspective precision. This proves \eqref{eq:archive_precisions}.

Using $b(r,\phi)=\phi(1+\phi)^{-1}[1-(1+qr)^{-1}]$ gives
\[
\Delta_{n,m}
=\frac{\phi}{1+\phi}
\left(\frac{1}{A+m\zeta}-\frac{1}{A+n\zeta}\right),
\]
which is \eqref{eq:access_closed}. Moreover,
\[
\zeta_\delta=\frac{\phi(1+\phi)}{(1+\delta+\phi)^2}>0,
\]
and direct differentiation yields
\begin{equation}\label{eq:feedback_derivative_proof}
\frac{\partial\Delta_{n,m}}{\partial\delta}
=\frac{\phi}{1+\phi}
\frac{(n-m)\zeta_\delta(A^2-nm\zeta^2)}
{(A+n\zeta)^2(A+m\zeta)^2}.
\end{equation}
At $m=0$ this is positive. For $0<m<n$, its sign is positive, zero, or negative according as $\zeta$ is below, equal to, or above $H=A/\sqrt{nm}$. Since $\zeta$ increases strictly from $1$ toward $1+\phi$, the three regimes follow, including the zero derivative at $H=1$, $\delta=0$. In the intermediate regime, solving $\zeta=H$ gives \eqref{eq:feedback_peak}. If $m=n$, the two marginal precisions are equal at every feedback level and the gap is zero. \qed

\subsection{Proof of Corollary~\ref{cor:access_calibration}}

An increase in $m$ strictly raises $r_U$ and leaves $r_F$ fixed. Monotonicity of $b$ implies that $\Delta_{n,m}$ falls. For a public-profile increment and $m<n$,
\[
\frac{\partial\Delta_{n,m}}{\partial\chi}
=b_\rho(r_F,\phi)-b_\rho(r_U,\phi)<0
\]
by strict concavity. The signs for prices, shares, and individual profits follow by differentiating \eqref{eq:prices} and \eqref{eq:profits} with respect to $\Delta$. Total industry profit is $\tau+\Delta^2/(9\tau)$, which strictly falls with a strict reduction of a positive gap. An archive opening preserves coverage because
\[
\bar v+b_U-c-\frac{3\tau}{2}+\frac{\Delta}{2}
=\bar v+\frac{b_F+b_U}{2}-c-\frac{3\tau}{2}
\]
increases, while the gap remains in $[0,3\tau)$. The same argument applies to public calibration. \qed

\subsection{Proof of Proposition~\ref{prop:allocation}}

Differentiating \eqref{eq:welfare_x} gives
\[
\frac{\partial W(x)}{\partial x}
=\Delta+\tau-2\tau x,
\qquad
\frac{\partial^2 W(x)}{\partial x^2}=-2\tau<0.
\]
The unconstrained maximizer is $1/2+\Delta/(2\tau)$. Since $\Delta\geq0$, only the upper constraint can bind, which gives \eqref{eq:efficient_share}. If $0\leq\Delta<\tau$, quadratic completion yields
\[
W(x_F^S)-W(x_F^*)
=\tau(x_F^S-x_F^*)^2
=\frac{\Delta^2}{9\tau}.
\]
If $\tau\leq\Delta<3\tau$, the planner chooses $x_F^S=1$. Subtracting \eqref{eq:welfare_equilibrium} from \eqref{eq:welfare_x} evaluated at one gives
\[
W(1)-W(x_F^*)
=-\frac{\tau}{4}+\frac{\Delta}{2}
-\frac{5\Delta^2}{36\tau}
=\frac{(3\tau-\Delta)(5\Delta-3\tau)}{36\tau}.
\]
For $\Delta>0$, the strict audience comparison follows immediately from \eqref{eq:share}. \qed

\subsection{Proof of Proposition~\ref{prop:performance}}

Integrating the reduction in quadratic prediction loss at the equilibrium cutoff gives \eqref{eq:forecasting}. Consumer surplus is
\[
CS^E=\bar v+Q^E-p_F^*x_F^*-p_U^*(1-x_F^*)
-\frac{\tau}{2}\bigl[(x_F^*)^2+(1-x_F^*)^2\bigr].
\]
Substitution of equilibrium prices gives \eqref{eq:cs}. Adding industry profit, or evaluating \eqref{eq:welfare_x} at $x_F^*$, gives \eqref{eq:welfare_equilibrium}. Differentiation of these three expressions, using $\Delta_s=b_{F,s}-b_{U,s}$, gives \eqref{eq:performance_weights}.

The covered interior condition implies $0\leq\Delta<3\tau$, hence $1/2\leq\omega_{CS}<2/3$. Consumer surplus is therefore strictly increasing under the stated assumptions. If $\Delta_s\geq0$, then $J_s=b_{U,s}+\omega_J\Delta_s>0$ for each of the three outcomes: if the gap does not increase strictly, both report values must increase. Thus a decrease requires $\Delta_s<0$.

When $\Delta\leq3\tau/2$, $\omega_Q\leq1$, so $Q_s$ is nonnegative. When $\Delta\leq9\tau/5$, $\omega_W\leq1$, so $W_s$ is nonnegative. Using \eqref{eq:share}, a decrease consequently requires $x_F^*>3/4$ and $x_F^*>4/5$, respectively. At the threshold a derivative can be zero if only $U$ improves; this does not affect the strict-loss statements.

For the public-profile formulas, $b_\rho(r,\phi)=\phi^2/(1+qr)^2$ gives \eqref{eq:t_ratio}. Thus
\[
J_\chi=b_{U,\chi}\{1-\omega_J(1-t_\chi)\},
\qquad J\in\{Q,W\}.
\]
The derivative is negative exactly when $t_\chi<1-1/\omega_J$, which yields \eqref{eq:calibration_forecast} and \eqref{eq:calibration_welfare}. \qed

\subsection{Proof of Proposition~\ref{prop:archive_welfare}}

Opening records leaves $b_F$ fixed and raises $b_U$ by $d$, so $\Delta_1=\Delta_0-d$. The coverage argument in the preceding corollary applies, and $0\leq\Delta_1<\Delta_0<3\tau$ ensures positive shares and markups throughout. Subtracting \eqref{eq:cs}, \eqref{eq:forecasting}, and \eqref{eq:welfare_equilibrium} at the two access levels gives \eqref{eq:archive_cs}--\eqref{eq:archive_w}. Since $\Delta_0+\Delta_1<6\tau$, the consumer-surplus change is strictly positive.

The prediction change is negative if and only if $\Delta_0+\Delta_1>3\tau$, and the surplus change is negative if and only if $\Delta_0+\Delta_1>18\tau/5$. The average endpoint share is
\[
\frac{x_{F,0}^*+x_{F,1}^*}{2}
=\frac12+\frac{\Delta_0+\Delta_1}{12\tau},
\]
which gives the two conditions in the proposition. Equalities give zero changes.

At complete access, $\Delta_1=0$ and $\Delta_0<3\tau$, so all three brackets are strictly positive. Applying this argument to every $m_0<n$ shows that complete opening maximizes gross surplus among the available access levels. \qed

\section{Overlapping historical states}\label{app:overlap}

The independent-date restriction permits source-by-source updating, but many archives contain reports about the same events. Suppose instead that all $m$ accessible rival reports concern $m$ of the $n$ dates already represented in the familiar history. There is one outcome signal per date, observed once. The remaining assumptions are unchanged. Write $r=\rho+\chi$ and $v_\delta=1/(1+\delta)$.

Conditional on the outcome signal, a pair of transformed reports has noise covariance
\begin{equation}\label{eq:overlap_noise}
\Omega=\phi I_2+\phi^2v_\delta\mathbf{1}\mathbf{1}^{\mathsf T}.
\end{equation}
The shared state produces the off-diagonal term. Define
\[
k=\frac{1}{\phi(1+\phi v_\delta)},\qquad
h=\frac{1+\phi v_\delta}{\phi(1+2\phi v_\delta)},\qquad
g=\frac{v_\delta}{1+2\phi v_\delta}.
\]
An unpaired familiar report contributes $k$ to its perspective precision. Inverting \eqref{eq:overlap_noise}, a paired observation contributes diagonal precision $h$ and off-diagonal precision $-g$. Independence across dates therefore gives the joint posterior precision matrix
\begin{equation}\label{eq:overlap_precision}
P_m=\begin{pmatrix}
r+(n-m)k+mh&-mg\\
-mg&r+mh
\end{pmatrix}
\equiv\begin{pmatrix}a&z\\z&e\end{pmatrix}.
\end{equation}
Its determinant $D=ae-z^2$ is positive. The \emph{marginal}, rather than conditional, perspective precisions relevant for purchasing one current report are
\begin{equation}\label{eq:overlap_marginals}
R_F=a-\frac{z^2}{e}=\frac{D}{e},\qquad
R_U=e-\frac{z^2}{a}=\frac{D}{a}.
\end{equation}
Current report values remain $b(R_F,\phi)$ and $b(R_U,\phi)$ because the new state is independent of the historical states. Correlation between perspectives does not change the value of a single report once its marginal precision is used.

Since $a-e=(n-m)k$, the familiar source is strictly better understood for $m<n$ and equally understood at complete access. Opening a paired rival report can also improve understanding of $F$: the rival report helps filter the shared historical state. This is why the source-by-source increments and the finite-change formulas in the main text cannot simply be reused.

The feedback reversal nevertheless survives. Take $\phi=1$, $\rho=0.1$, $\chi=0$, and $n=5$. At $\delta=1$, differentiating the gap obtained from \eqref{eq:overlap_marginals} gives approximately $0.0089773$ when $m=0$ and $-0.0015871$ when $m=1$. Access to one rival report reverses the sign. The exact threshold is different: at $m=1$ and $\delta=0$, the derivative is approximately $0.0059260$ with overlapping dates, whereas it is negative with independent dates. The qualification attached to Proposition~\ref{prop:access_feedback} is therefore economically consequential.

Table~\ref{tab:overlap} repeats the archive comparison in Figure~\ref{fig:archive}, now with overlapping dates. Opening one record improves both report values but lowers aggregate forecasting benefit and gross surplus. Complete opening raises all three outcomes.

\begin{table}[htbp]
\centering
\caption{An archive of reports about shared states}\label{tab:overlap}
\small
\begin{tabular}{lrrrrrr}
\toprule
$m$ & $b_F$ & $b_U$ & $x_F^*$ & $Q^E$ & $CS^E$ & $W^E$\\
\midrule
0 & 0.4194 & 0.0833 & 0.9480 & 0.4019 & 2.1202 & 2.3456\\
1 & 0.4199 & 0.2962 & 0.6650 & 0.3785 & 2.2052 & 2.3438\\
5 & 0.4200 & 0.4200 & 0.5000 & 0.4200 & 2.2637 & 2.3887\\
\bottomrule
\end{tabular}
\par\smallskip
\begin{minipage}{0.95\textwidth}
\footnotesize
Notes: The parameters are those of Figure~\ref{fig:archive}. Each rival record now concerns a date in the familiar history. All listed equilibria satisfy coverage and interiority. The numerical entries use joint filtering in \eqref{eq:overlap_precision}--\eqref{eq:overlap_marginals}.
\end{minipage}
\end{table}

The benefit of complete access extends beyond this example. Additional observations weakly reduce each marginal posterior variance. At $m=n$, both reports consequently have a common value $b^*$ at least as large as the familiar report's value at any earlier access level. From an incomplete covered interior equilibrium, aggregate forecasting benefit was strictly below that earlier familiar value, since a positive share bought the less informative rival report. At complete access it equals $b^*$. Horizontal mismatch is also minimized at the symmetric allocation, so gross surplus strictly rises. Consumer surplus rises as well: both report values weakly increase, at least one strictly, and the segment between the initial values and $(b^*,b^*)$ lies in the covered interior region, where Proposition~\ref{prop:performance} applies. This argument establishes the complete-opening comparison without claiming that every intermediate access level improves forecasting or gross surplus.

\end{document}